\pdfoutput=1
\documentclass[12pt,letterpaper,amsmath]{article}

\usepackage{amsfonts,amsmath,amssymb,mathrsfs}
\usepackage{graphicx}

\begin{document}

\def\beq{\begin{equation}}
\def\eeq{\end{equation}}
\def\bea{\begin{eqnarray}}
\def\eea{\end{eqnarray}}
\def\nn{\nonumber}
\def\ts{}

\def\a{\alpha}
\def\b{\beta}
\def\r{\rho}
\def\s{\sigma}
\def\m{\mu}
\def\n{\nu}
\def\k{\kappa}
\def\g{\gamma}
\def\th{\theta}
\def\L{\Lambda}
\def\D{\Delta}
\def\la{\langle}
\def\ra{\rangle}
\def\o{\omega}
\def\d{\delta}
\def\p{\partial}
\def\Se{$S_E$ }
\def\Sa{$S_{\rm atmo}$ }

\def\tb{\bar{t}}
\def\tt{\tilde{t}}
\def\phib{\bar{\phi}}
\def\phit{\tilde{\phi}}
\def\tu{\tilde{u}}
\def\hv{\hat{v}}
\def\thh{\hat{\theta}}
\def\phih{\hat{\phi}}

\def\half{\textstyle{\frac{1}{2}}}
\def\quarter{\textstyle{\frac{1}{4}}}

\begin{center} {\LARGE \bf
Slices of the Kerr ergosurface}
\end{center}

\vskip 5mm
\begin{center} \large
Ted Jacobson\footnote{E-mail: jacobson@umd.edu} and
Yee-Ann Soong\footnote{E-mail: soongy@umd.edu}
\end{center}


\vskip  0.5 cm 
\begin{center}{}{\it
Center for Fundamental Physics\\
Department of Physics\\
University of Maryland\\
College Park, MD 20742-4111, USA}
\end{center}

\vskip 1cm

\begin{abstract}
The intrinsic geometry of the Kerr ergosurface
on constant Boyer-Lindquist (BL), Kerr,
and Doran time slices is characterized.
Unlike the BL slice, which had been previously 
studied, the other slices (i) do not have conical singularities
at the poles (except the Doran slice in the
extremal limit), (ii) have finite polar circumference 
in the extremal limit, and (iii)  for sufficiently large spin parameter
fail to be isometrically embeddable as a surface of revolution
above some latitude. The Doran slice develops an embeddable polar cap
for spin parameters greater than about 0.96.

\end{abstract}

\newpage
\section{Introduction}

The ergosurface of the Kerr spacetime is the locus at
which the asymptotic time translation symmetry becomes
lightlike. This symmetry is spacelike inside the
ergosurface, so negative energy states of matter are
allowed there~\cite{Penrose:1969pc}. 
The existence of these negative
energy states is probably central to the mechanism
by which rotational energy is extracted from
astrophysical black holes~\cite{Begelman:1984mw, Meier01}.
The Penrose
process~\cite{Penrose:1969pc,Penrose:1971uk} 
is an idealized example of such a
mechanism, whose discovery led directly to the
realization that black holes behave as thermodynamic
systems in equilibrium, with an entropy proportional
to the surface area of the event horizon and a
temperature proportional to the surface gravity~\cite{Bekenstein:1973ur}.

Given the importance of the ergosurface, it is natural
to examine its intrinsic two-dimensional (2d)
geometry. While this geometry plays no direct role
in dynamical processes, it could be useful 
for physical intuition and visualization,
and is in any case mathematically interesting. Unlike
the event horizon, however, the ergosurface is not a
null surface, and therefore different spacelike slices
of it determine different 2d geometries. In this paper
we examine and compare the geometry of three different
slices, those defined by the  Boyer-Lindquist (BL),
Kerr, and Doran~\cite{Doran:1999gb} time coordinates.

The geometry of the BL slice has already been extensively
studied~\cite{Sharp81,Kokkotas:1988fx,Pelavas:2000za}. 
Peculiar features of that geometry are that
the poles have conical singularities, and that in
the limit of maximal rotation (extreme Kerr geometry)
the surface becomes infinitely long in the polar
direction. This happens because the ergosurface
coincides with the horizon at the poles, and the
horizon recedes to infinite distance on a surface of
constant BL time coordinate in the extremal limit. By
contrast, the horizon remains at a finite separation
on constant Kerr or Doran time surfaces, since these 
surfaces are defined by geodesics freely falling across 
the horizon.

The remainder of this paper is organized as follows.
In section 2 we present the Kerr metric in the three
different coordinate systems and obtain the 2d
metrics for the corresponding ergosurface slices.
Section 3 discusses general properties of axisymmetric
2d geometries, and section 4 presents the results for
the slices of the Kerr ergosurface.
Mathematica was used for most of the computations.

\section{Kerr spacetime and the ergosurface}

The Kerr black hole spacetime has both time translation symmetry
and axial rotation symmetry. 
The BL, Kerr, and Doran coordinate systems
all have in common 
two coordinates, called $r$ and $\theta$,
which are constant in these symmetry
directions. The remaining two 
coordinates are $(t,\phi)$ for BL, 
$(\tt,\phit)$ for Kerr ($\tt$ is commonly called
$v$), and 
$(\tb,\phib)$ for Doran.
The ``time" coordinates $t$, $\tt$, and $\tb$
differ from each other only by the addition of a function
of $r$, as do the azimuthal angle coordinates
$\phi$, $\phit$, and $\phib$. 
The relation between the time coordinates is given by 
\beq\label{t}
dt+\ts{\frac{\b}{1-\b^2}}\, dr =d\tb = d\tt -\ts{\frac{1}{1+\b}}\, dr
\eeq
and the angles are related by 
\beq\label{phi}
d\phi+\frac{\b}{1-\b^2}\frac{a}{r^2+a^2}\, dr
=d\phib=d\phit-\frac{1}{1+\b}\frac{a}{r^2+a^2}\, dr
\eeq
with 
\beq\label{beta}
\b^2=2Mr/(r^2+a^2),
\eeq
where $M$ and $a$ are the mass and spin parameter
(angular momentum divided by mass) of the 
Kerr spacetime, and $\b$ is the positive root of (\ref{beta}).

The line element 
in these three coordinate systems  takes the form
\bea
ds^2&=&-(1-\a^2)\, dt^2
+\a^{-2}\b^2(1-\b^2)^{-1}\, dr^2
-2 \a^2 a\sin^2\!\theta\, dt d\phi\nonumber\\
&&\qquad\qquad+\r^2\, d\theta^2
+(r^2+a^2+\a^2 a^2 \sin^2\!\theta)\sin^2\!\theta\, d \phi^2\label{BL}\\
\nn\\
&\vspace{4cm}=& -(1 -\a^2)\, d\tt^2 
+ 2(d\tt- a \sin^2\!\theta\, d\phit)dr
- 2 \a^2 a \sin^2\!\theta\, d\tt d\phit \nn \\
&&\qquad\qquad  + \rho^2\, d\theta^2 
+( r^2 +a^2+\a^2 a^2 \sin^2\!\theta)\sin^2\!\theta\,  d\phit^2\label{EF}\\
\nn\\
&=&-d\tb^2 
+ \left(\a^{-1}\b\,  dr 
+ \a( d\tb - a\sin^2\!\theta \, d\phib) \right)^2 \nn \\
&&\qquad\qquad + \rho^2\, d\theta^2 + (r^2 +a^2) \sin^2\!\theta \, d\phib^2,
\label{Doran}
\eea
where 
\beq\label{alpharho}
\a^2=2Mr/\r^2\qquad\mbox{and}\qquad \r^2 = r^2 + a^2\cos^2\!\theta,
\eeq
and $\a$ is the positive root.
Note that, since the coordinates differ only by functions of $r$, 
the coefficients of all terms not involving $dr$ are
equal for the three coordinate systems.

We now summarize some key properties of the Kerr 
geometry in these coordinates.
In BL coordinates the metric has just one off-diagonal ($dt d\phi$)
term,
and the constant $t$ surfaces are orthogonal to 
stationary, zero angular momentum observers (ZAMO's).
In Kerr coordinates
(which are sometimes called advanced Eddington-Finkelstein 
coordinates by analogy with the non-rotating, Schwarzschild case) 
the curves 
$d\th=d\phit=d\tt=0$ are lightlike geodesics with
zero angular momentum and affine parameter $r$. The surfaces
of constant $\tt$ are {\it timelike}, 
so $\tt$ is actually a ``space coordinate" rather than a time coordinate.
(In the non-rotating case it is lightlike.)
In Doran coordinates,
the curves  $d\th=d\phib=\b dr+\a^2d\tb=0$ are 
timelike geodesics with zero angular momentum, unit energy
(at rest at infinity), and proper
time $\tb$. The surfaces
of constant $\tb$ are orthogonal to
these geodesics.

The event horizon of the black hole is located where
$\b=1$, i.e.
\beq
r_h=M+(M^2-a^2)^{1/2}.
\eeq
The BL $t$ coordinate
is singular at the horizon, where it runs to infinity.
This can be seen in the coordinate transformation
(\ref{t}). The relation between $\tb$ and $\tt$ on the other hand is finite
in this limit. Similar relations hold between the various 
angle coordinates. 

The ergosurface occurs where the time translation 
symmetry becomes lightlike, at $\a=1$, i.e. 
\beq\label{re}
r_e(\th)=M+(M^2-a^2 \cos^2\!\th)^{1/2}.
\eeq
This 
surface coincides with the
horizon at the poles, but lies outside
the horizon at all other latitudes. 
On this surface $dr$ is related to $d\th$ via
\beq\label{dr}
dr=\frac{dr_e}{d\th}\, d\th = \frac{a^2\cos\th \sin\th}{(M^2-a^2 \cos^2\!\th)^{1/2}}\, d\th.
\eeq
We are interested in the 
2d intersection of the ergosurface with a surface of constant 
$t$, $\tt$, or $\tb$. 
The line element restricted to these 2d surfaces is
\bea
\mbox{(BL)}&&\hspace{-5mm}2r_e\, d\theta^2
+2(r_e+a^2 \sin^2\!\theta)\sin^2\!\theta\, d \phi^2+\b^2(1-\b^2)^{-1}\, dr^2
\label{BL2}\\ \nn\\
\mbox{(Kerr)}&&\hspace{-5mm}2r_e\, d\theta^2 
+2(r_e+a^2 \sin^2\!\theta)\sin^2\!\theta\,  d\phit^2-2a \sin^2\!\theta\, dr d\phit
 \label{EF2}\\ \nn\\
\protect\hspace{-2cm}
\mbox{(Doran)}&&\hspace{-5mm}2r_e\, d\theta^2 + 2(r_e+a^2 \sin^2\!\theta) \sin^2\!\theta \, d\phib^2+
\b^2\,  dr^2
- 2\b a\sin^2\!\theta \, dr d\phib,
\label{Doran2}
\eea
where $dr$ is given by (\ref{dr}), and we have adopted units with $M=1$.
In the limiting case $a=0$, the ergosurface coincides with the 
horizon, which sits at fixed radius. All three slices then coincide, and the 
geometry is just a 2-sphere of radius $r_h=2M$. 

\section{Axisymmetric 2d geometry}
\label{Axigeom}
The three 2d metrics (\ref{BL2}-\ref{Doran2}) 
all take the form
\beq \label{2dgen}
A\, d\th^2 + 2B\, d\th d\phi + R^2\, d\phi^2
\eeq
where $A$, $B$, and $R$ are functions only of $\th$,
and are symmetric under reflection about the equator 
$\th=\pi/2$.
$R$ is the circumferential radius of the symmetry 
circles, i.e.\ the lines of latitude (constant $\th$), and is the same in 
all three cases, 
\beq\label{R}
R= \left(2r_e(\th)+2a^2 \sin^2\!\theta\right)^{1/2}\sin\theta.
\eeq
Changing coordinates via
\bea\label{phihat}
d\phih=d\phi +(B/R^2)\, d\th, \qquad du=(A-B^2/R^2)^{1/2} d\th
\eea
puts (\ref{2dgen}) into the standard form
\beq \label{2dspec}
du^2  + R^2\, d\phih^2.
\eeq
The intrinsic geometry is fully described by the
function $R(\th(u))$, which gives the circumferential 
radius as a function of perpendicular distance 
$u$ from the pole. 
The range of $u$ (unlike $\theta$) represents a 
geometrically intrinsic property, and is given by 
the integral of $du$ over $\theta$ from $0$ to $\pi$.

The Gaussian curvature $k$ of the surface
provides a purely local handle on the geometry.
For the metric (\ref{2dspec}) it is 
given by 
\beq\label{kspec}
k=-R_{,uu}/R
\eeq
where the comma denotes derivative with respect to the
following variable. This may be
expressed in terms of derivatives with respect to $\th$ using
the chain rule, which yields
\beq\label{kgen}
k=-\frac{R_{,\th\th}}{R(A-B^2/R^2)} +\frac{R_{,\th}(A-B^2/R^2)_{,\th}}{2R(A-B^2/R^2)^{2}}.
\eeq
Note that since the radius $R$ goes to zero at the poles,
the curvature
(\ref{kspec}) diverges there unless $R_{,uu}$
approaches zero
at least as fast as $R$. 

Even if the curvature
remains finite as the pole is approached, 
there is a conical singularity at $\th=0$ unless 
$R_{,u}\rightarrow1$.  Indeed, if $R=\s u + O(u^2)$,
one can absorb the constant factor $\s$ into 
a rescaled angle $\s\phih$, which 
then ranges from
$0$ to $2\pi\s$. The deficit angle is thus 
given by $\d=2\pi(1-\s)$, where 
$\s=R_{,u}(u=0)$. In 
terms of the original metric coefficients, this
may be expressed as
\beq\label{deficit1}
\d=2\pi\left(1-\frac{R_{,\th}}{(A-B^2/R^2)^{1/2}}\right)_{\th\rightarrow0}.
\eeq
In the BL case $B=0$, and in the other two cases $B$ vanishes
fast enough with $\th$ that it does not contribute to (\ref{deficit1}),
hence for the cases at hand we have the simpler formula
\beq\label{deficit}
\d=2\pi\left(1-R_{,\th}/A^{1/2}\right)_{\th\rightarrow0}.
\eeq

The surface is isometrically embeddable 
as a surface of revolution (hereafter just 
``embeddable") in 3d Euclidean space
with cylindrical coordinates $(R,\phih,z)$
if and only if there exists a real height 
function $z(R)$ such that 
\beq\label{embed}
du^2= dR^2 + d(z(R))^2,
\eeq
where $d(z(R))=z_{,R}\, dR$.
Such a function exists if and only if $dR\le du$, 
i.e.\ $R_{,u}\le1$.\footnote{If $R_{,u}>1$ then (\ref{embed}) implies
$dz^2<0$, so $dz$ is pure imaginary. One can then identify $|dz|$ 
as the change of a time coordinate in a 3d Minkowski space,
thus determining a Minkowskian embedding, as was done
in Ref.~\cite{Sharp81}.}
That is, if the surface is to be embeddable then
$R$ must not grow faster than $u$.\footnote{If 
the condition that the embedding be
a surface of revolution is dropped then, according to
the Janet-Cartan theorem, any analytic 2d metric can
be locally isometrically embedded in 3d Euclidean space~\cite{book}.}
When it exists, $z(R(u))$ is determined by the differential equation
\beq\label{dzdu}
dz/du=\left(1-(R_{,u})^2\right)^{1/2}.
\eeq
In terms of the original coordinate $\th$ and the  
original metric functions this becomes\footnote{Ref.~\cite{Sharp81} 
states that no axisymmetric embedding exists
if $B\ne0$. This is evidently untrue as long as one
allows for a $\th$-dependent shift (\ref{phihat}) of $\phih$ 
relative to $\phi$.}
\beq
\frac{dz}{d\th} = \left(A-B^2/R^2 - (R_{,\th})^2\right)^{1/2}.
\eeq
Note that the embeddability condition $R_{,u}\le1$ is
independent of the sign of the curvature (\ref{kspec}), 
i.e.\ the sign of $-R_{,uu}$. 

\section{Geometries of Kerr ergosurface slices}

%
\begin{figure}[ph]
\begin{center}
\includegraphics[width=4in]{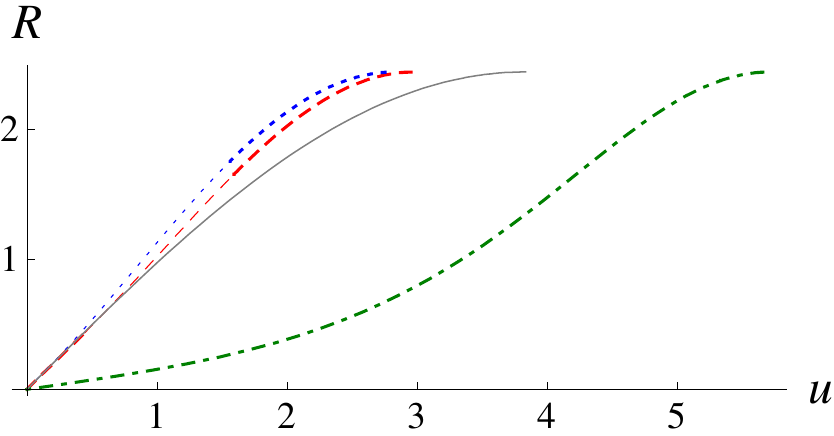}
\includegraphics[width=4in]{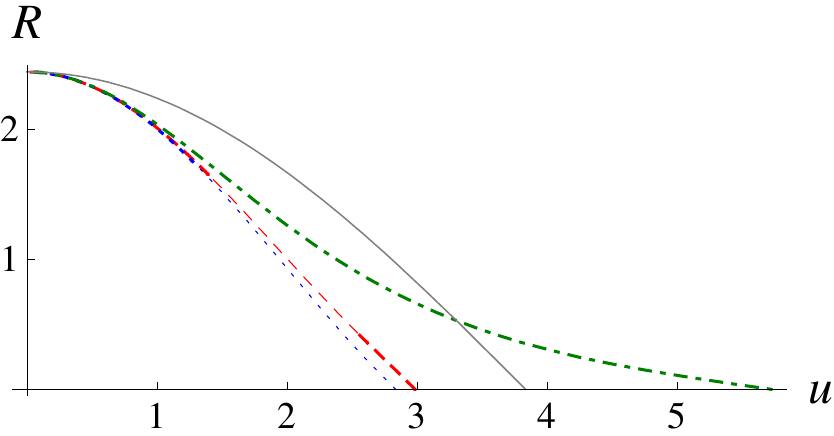}
\end{center}\caption{\small
Circumferential radius $R$ of latitude circles 
vs.\  
longitudinal
distance $u$ for $a=0.99$. 
In the upper graph the range of $u$  
runs from $0$ at the pole to its value at the
equator (which is different for each slice).
In the lower graph $u$ runs in the opposite
direction, with $u=0$ at the equator rather 
than at the pole.
The BL curve is dot-dashed green, the
Kerr curve is dotted blue, and the Doran
curve is dashed red. 
For comparison, the solid grey line displays the
graph for a round sphere with the same equatorial circumference. 
The curves 
(other than that for the sphere)
are thick where
the surface is embeddable (i.e.\ where the magnitude of
the slope is less than
unity), and thin where it is not.
The sign of the Gaussian curvature $k$ at each $u$ 
is opposite to the curvature of the graph, according to
(\ref{kspec}). 
%
%
\label{Rvsu} 
}
\end{figure}

As stated above, in the non-rotating limit the BL, Kerr, and Doran slices 
of the ergosurface all coincide with a 2-sphere
of radius $2M$. As $a$ grows, the slices deviate increasingly from
each other, 
becoming maximally different in the extremal case
$a=M$. (We do not consider trans-extremal cases here.)
The Kerr and Doran $R(u)$ functions are
extremely close to each other until $a$ is very close to unity,
e.g.\ for $a=0.7M$ they differ by less than one percent,
while the BL slice deviates significantly for smaller values of 
$a$. In the remainder of this
section we adopt units with $M=1$.

A representative example is shown in Fig.~\ref{Rvsu}
which displays 
numerically generated plots of $R$ (circumference/$2\pi$) 
vs. $u$ (longitudinal distance) 
for the highly
spinning case $a=0.99$. The content of these two
figures is the same, except the different cases are aligned at the pole in
the upper figure and at the equator in the lower figure, to
better compare the geometries.  
The legend scheme is explained in the figure caption,
and is maintained for all the following plots.

The BL slice (dot-dashed green) curvature is 
positive below some latitude and negative above.  
The distance from equator to pole is much longer than
for the other slices, and in fact grows without bound as $a$ approaches unity.
There is a conical deficit at the pole,
indicated by the slope of $R(u)$ which is less than unity at $u=0$.
In fact there is a conical singularity on the BL slice for all nonzero values
of $a$, with deficit angle (\ref{deficit}) given by $2\pi(1-(1-a^2)^{1/2})$. 
The BL slice is globally embeddable for all values of $a$.
(These results for the BL case were previously obtained in 
Refs.~\cite{Sharp81,Kokkotas:1988fx,Pelavas:2000za}.)

The Kerr slice (dotted blue)  curvature is also 
positive below some latitude and negative above. 
The distance from equator to pole remains finite as
$a$ approaches unity. The slope of $R(u)$ at $u=0$ 
is unity so the geometry is 
smooth at the pole, i.e.\ there is no conical deficit,
as can be easily verified directly using (\ref{deficit})
for all values of $a$.
The curvature is negative at the pole, where the slope is
increasing from unity ($R_{,uu}>0$). When the slope begins decreasing 
the curvature becomes positive, but the surface remains non-embeddable
until the slope drops down below unity, 
which for the example shown occurs around $u\approx1.5$.

The Doran slice (dashed red) curvature 
is positive at the equator, becomes negative
above some latitude, and then becomes 
positive again near the pole.
Like for the Kerr slice,
the distance to the pole remains finite for $a=1$, 
and the geometry is smooth at the pole (except for $a=1$, as discussed below).
The example shown in Fig.~\ref{Rvsu} is embeddable
in a polar cap region $u\lesssim0.4$ as well as for 
an equatorial region $u\gtrsim1.5$. 
(An embeddable polar region
occurs for $a$ greater than a value between
$0.95$ and $0.96$.)
Moving from the equator to the pole,
the curvature turns negative
roughly halfway through the non-embeddable
region and becomes positive again roughly  halfway 
through the embeddable polar cap. 

\subsection{Curvature}

The curvature at the equator
is the same on all three slices,
and equal to $k_{eq}=(4+5a^2)/(16+8a^2)$. The
non-spinning case has $k_{eq}=1/4$ and the extremal case
has $k_{eq}=3/8$. To see why it is the same on all the slices, note that
the geometry is reflection symmetric about the
equatorial plane, so $dr$ vanishes there; in fact, as (\ref{dr}) shows,
it vanishes as $\cos\th$. 
Thus the metric components on the equator 
agree for all three slices. Moreover, 
reflection symmetry implies $R_{,\th}=0$,
so the second term of (\ref{kgen}) vanishes,
and the first term of (\ref{kgen}) agrees since $A$ and $B$
agree on the equator and $R$ agrees everywhere.

\begin{figure}[bhp]
\begin{center}
\includegraphics[width=4in]{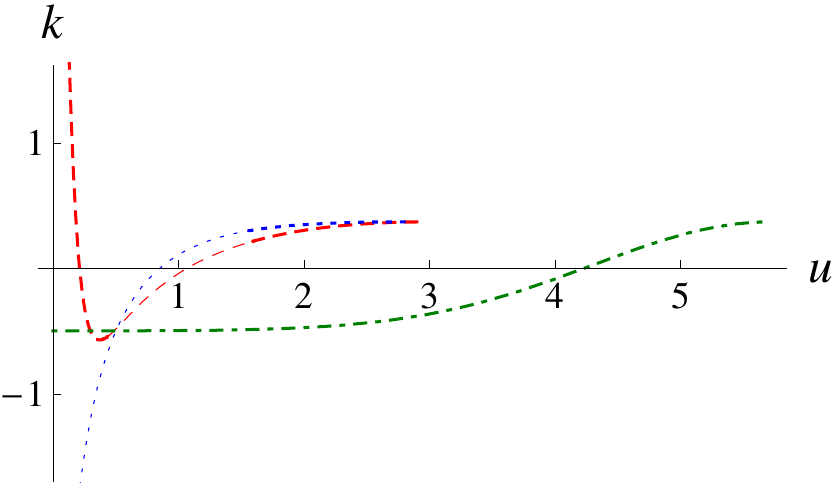}
\end{center}\caption{\small
Curvature vs.\ polar distance, 
$a=0.99$. 
The curvature is positive at the equator,
and becomes negative at a higher latitude.
On the Doran slice it turns around and goes to
positive values near the pole. 
As in the previous figures, the curves are thick where 
the surface is embeddable. Note that the sign of the curvature
does not determine embeddability.
\label{kvsu}}
\end{figure}
The curvature as a function of polar distance is 
shown in Fig.~\ref{kvsu} for the representative case
$a=0.99$. 
For all three slices the curvature is positive at the equator,
and becomes negative at a higher latitude. It levels
out as the pole is approached on the BL slice,
keeps dipping to lower values on the Kerr slice,
and on the Doran slice it turns around and goes to
positive values near the pole. 

\begin{figure}[tbhp]
\begin{center}
\includegraphics[width=4in]{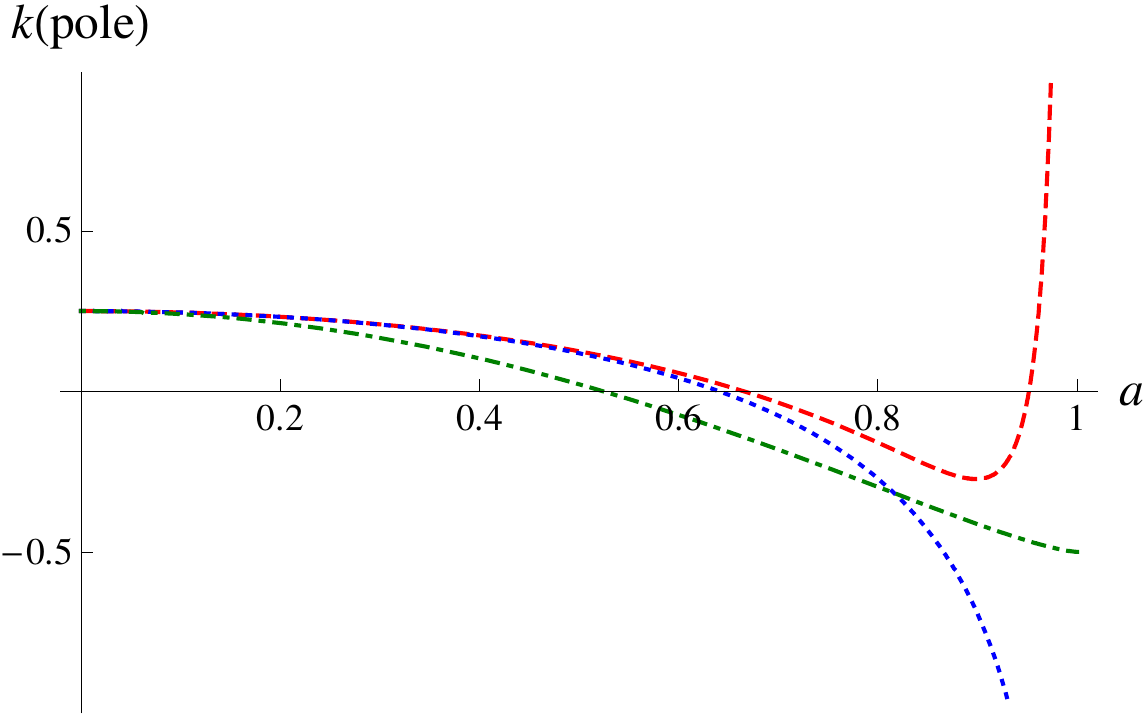}
\end{center}\caption{\small
Curvature at the pole vs. spin parameter $a$.
All three slices have 
positive curvature at the pole
for sufficiently small values of 
$a$, and they first develop negative curvature
for $a$ approximately equal to $0.52$, $0.64$, and $0.67$ 
in the BL, Kerr, and Doran cases respectively.
They have rather different
behavior as $a$ grows close to 1.
The polar curvature goes to $-1/2$ in the BL case,
and diverges negatively in the Kerr case. In the Doran 
case it becomes positive again for $a$ greater than approximately 
$0.95$,  and diverges positively as $a$ approaches 1. 
A polar region is embeddable in the Kerr and Doran cases 
if and only if the curvature at the pole is positive.   
\label{kpole}}
\end{figure}

The curvature at the pole itself is shown
as a function of $a$ in 
Fig.~\ref{kpole}. 
In the BL case this is the limiting curvature
as the conical singularity at the pole is approached.
In the Kerr and Doran cases, the geometry at the pole 
is smooth so, as explained in section~\ref{Axigeom}, 
$R_{,u}=1$ there. Thus for positive curvature ($-R_{,uu}>0$)
at the pole, a polar region with $R_{,u}<1$ exists, which
is embeddable, while for negative
curvature it is not embeddable at the pole. 

In the precisely extremal case $a=1$, the curvature 
approaching the pole diverges negatively as 
$-u^{-1}$ on both the 
Kerr and Doran slices.
This appears at first to contradict 
Figs.~\ref{kvsu} and \ref{kpole} which show a {\it positive}
divergence of the curvature in the Doran case. 
The resolution of this apparent contradiction is that, 
as $a\rightarrow1$, the polar region of positive curvature
contracts to an infinitesimal region including only the pole 
itself. The asymptotic behavior $-u^{-1}$ describes an infinite
negative dip before this infinitely narrow, infinitely positive spike.

 \subsection{Embedding diagrams}

\begin{figure}[ptbh]
\begin{center}
\includegraphics[width=4in]{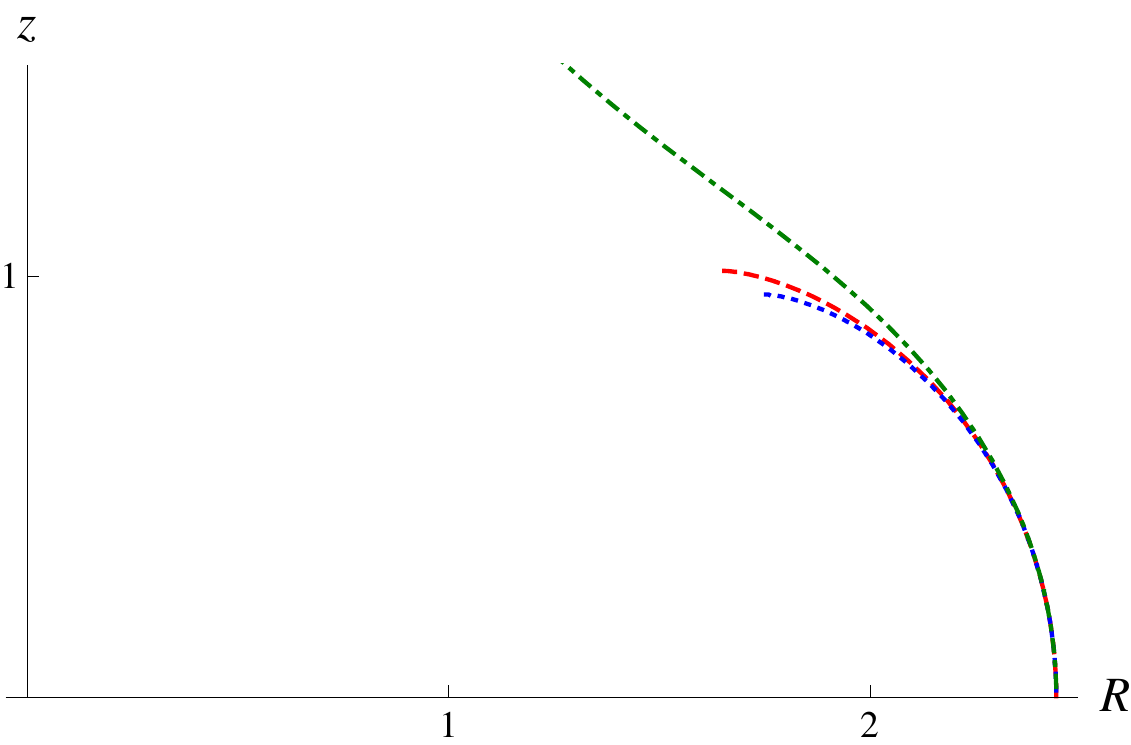}
\end{center}\caption{\small
Embedding of equatorial region, $a=0.99$.
The surface of revolution is obtained by rotating the graph
about the vertical, $z$ axis, and reflecting it across the $R$ axis.
The curve for the
BL slice is truncated  at $z=1.5$. When continued, 
it tapers to a conical singularity at the axis. 
The Kerr and Doran slices have rather similar embeddings.
As implied by (\ref{dzdu}), 
the slope $dz/dR$ vanishes at the point where the
embedding fails to exist.
\label{eqembed}}
\end{figure}

\begin{figure}[ptbh]
\begin{center}
\includegraphics[width=4in]{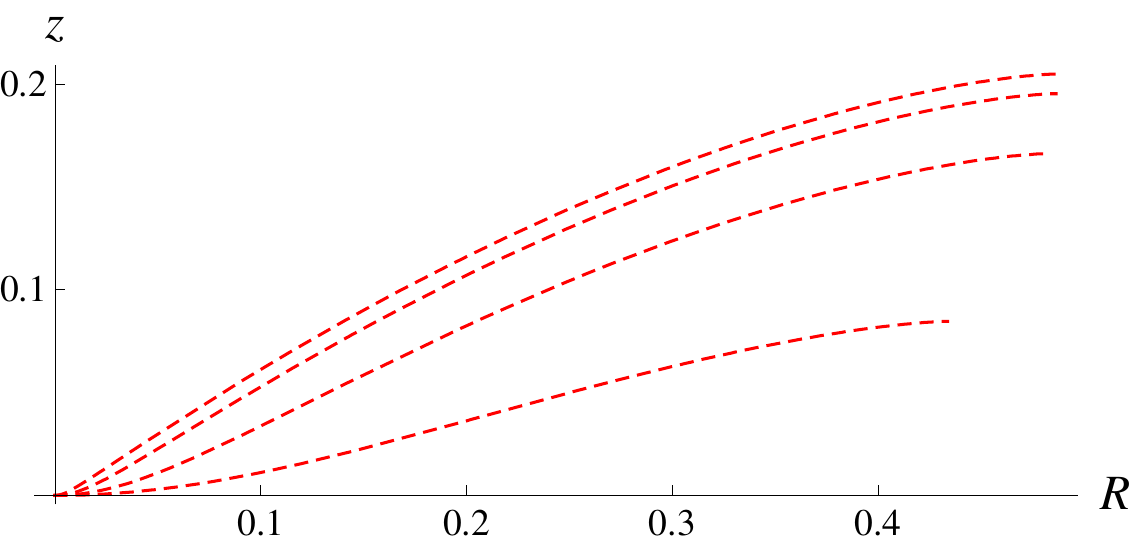}
\end{center}\caption{\small
Embedding of Doran polar region,  $a=0.99$, 0.999, 0.9999, 0.99999.
In the limiting case $a=1$ a conical singularity develops at the
pole.
\label{capembed} }
\end{figure}

An embedding diagram for the region including the equator is
shown in Fig.~\ref{eqembed} for the case $a=0.99$.
An embedding diagram for the Doran polar cap is shown separately
in Fig.~\ref{capembed}, for a sequence of increasing values of
$a$. 
As $a$ approaches closer to unity, the embedding
is more sharply curved at the pole. In the 
limiting case $a=1$ it develops a conical singularity at the
pole. The reason is that, rather than vanishing
at the pole, $dr$ is given by $a\, d\th$, since the denominator
of (\ref{dr}) vanishes as $\sin\th$, cancelling the $\sin\th$
in the numerator. The deficit angle, calculated from 
(\ref{deficit}), is found to be $2\pi(1-\sqrt{2/3})$. 

The polar cap could be placed in a common embedding diagram with the
the equatorial region, faithfully showing the $R$ values relative to
a common axis. However, the relative position in the $z$ direction
would have no significance since these two patches are not
joined by an embedding.
Also, the sign of $z$ has no effect on the induced geometry
of the embedded surface, so the polar cap embedding
could just as well be displayed reflected across the
$R$ axis, as a bump rather than as a dimple
at the pole. We have chosen to display it as a dimple in
Fig.~\ref{capembed} just because,
if  the radial distance in the embedding space is
(incorrectly) identified with this $r$ coordinate,
the increase of the $r$ coordinate of the ergosurface
(\ref{re}) with $\th$ is mimicked.

\section{Summary}

We have analyzed and compared the geometries
of three different spatial slices of the ergosurface 
of the Kerr spacetime, corresponding to constant
values of Boyer-Lindquist, Kerr and Doran time coordinates.
We found that, unlike the BL slice which has a conical singularity,
both the Kerr and Doran slices are smooth at the pole,
except for the extremal case of the Doran slice.
Also unlike the BL slice, the distance from equator to
pole remains finite as the spin parameter $a$ 
approaches the extremal value $M$. All the slices develop
negative curvature above some latitude for sufficiently high spin,
and the Doran slice develops also a positive curvature polar
cap above $a\approx 0.95 M$. In the extremal limit
$a=M$ this becomes a conical singularity at the pole, 
with a deficit angle $2\pi(1-\sqrt{2/3})$. 

The slices differ in where they can be
isometrically embedded as surfaces of revolution 
in a 3d Euclidean space. The BL slice is fully embeddable
for all spin parameters, and the Kerr and Doran slices
fail to be embeddable above some latitude when the
spin parameter is sufficiently large. For spin parameter
$a\gtrsim 0.96M$ the Doran slice is also embeddable at
high latitudes, in a polar cap. 
We showed some embedding diagrams, and 
indicated the embeddable regions on plots
of circumference vs. radius. These plots allow easy
comparison of the geometries of the three slices.

\section*{Acknowledgments}
This work was supported in part by the National
Science Foundation under grant PHY-0601800.

\renewcommand{\theequation}{A.\arabic{equation}}
  \setcounter{equation}{0}  

\end{document}